\documentclass[preprint,prc,showpacs,preprintnumbers,amsmath,amssymb,floatfix]{revtex4}
\usepackage{hyperref}
\usepackage{booktabs}
\usepackage{threeparttable}
\usepackage{amsmath}
\usepackage{graphicx}% Include figure files
\usepackage{dcolumn}% Align table columns on decimal point
\usepackage{bm}% bold math
\usepackage{ulem} %% for strike-through
\usepackage[usenames]{color}
\usepackage{epstopdf}
\usepackage{epsfig}
\usepackage{float}
\usepackage{subfigure}
\usepackage{multirow}
\usepackage[version=3]{mhchem}

\newcommand{\nc}{\newcommand}       % new command
\newcommand{\renc}{\renewcommand}   % re-new command
\nc{\vc}[1] {\mbox{\boldmath $#1$}} % boldmath(vector)
\nc{\del}       {\partial}              % bra state
\nc{\bra}       {\langle}               % bra state
\nc{\ket}       {\rangle}               % ket state
\nc{\bras}[1]   {\langle #1|}           % bra state
\nc{\kets}[1]   {|#1\rangle}            % ket state
\nc{\mapleft}[1]{           % something under arrow
	\smash{\mathop{\,          %
			\hbox to 1.5cm{\rightarrowfill}\, }\limits_{#1}}}
\nc{\beq}     {\begin{eqnarray}} \nc{\eeq}    {\end{eqnarray}}
\nc{\nn}      {\\\nonumber} \nc{\vs}      {\vspace{-0.275cm}}
\nc{\fra}    {\frac{1}{2}}
\nc{\mb}        {\mathbf}
\nc{\wt} {\widetilde}
\renc\thesubsection{\arabic{subsection}}

\begin{document}
	
\preprint{}

\title{Effects of symmetry energy on the radius and tidal deformability of neutron stars in relativistic mean-field model}

%%%% To generate auto affiliation numbers please use \author{}\affil{} command

\author{Jinniu Hu}
\affiliation{School of Physics, Nankai University, Tianjin 300071,  China\\
	Strangeness Nuclear Physics Laboratory, RIKEN Nishina Center, Wako, 351-0198, Japan \email{hujinniu@nankai.edu.cn}}
\email{hujinniu@nankai.edu.cn}

\author{Shishao Bao}
\affiliation{School of Physics and Information Engineering, Shanxi Normal University, Linfen 041004, China}

\author{Ying Zhang}%%% Use optional bracket [3] to change the respective address
\affiliation{Department of Physics, Faculty of Science, Tianjin University, Tianjin 300072, China\\
	Strangeness Nuclear Physics Laboratory, RIKEN Nishina Center, Wako, 351-0198, Japan}

\author{Ken'ichiro Nakazato}
\affiliation{Faculty of Arts and Science, Kyushu University, 744 Motooka, Nishi-ku, Fukuoka 819-0395, Japan}
\author{Kohsuke Sumiyoshi}
\affiliation{National Institute of Technology, Numazu College, Ooka 3600, Numazu, Shizuoka 410-8501, Japan}
\author{Hong Shen}
\affiliation{School of Physics, Nankai University, Tianjin 300071,  China}

\date{\today}

\begin{abstract}%
The radii and tidal deformabilities of neutron stars are investigated in the framework of relativistic mean-field (RMF) model with different density-dependent behaviors of symmetry energy. To study the effects of symmetry energy on the properties of neutron stars, an $\omega$ meson and $\rho$ meson coupling term is included in a popular RMF Lagrangian, i.e. the TM1 parameter set, which is used for the widely used supernova equation of state (EoS) table. The coupling constants relevant to the vector-isovector meson, $\rho$, are refitted by a fixed symmetry energy at subsaturation density and its slope at saturation density, while other coupling constants remain the same as the original ones in TM1 so as to update the supernova EoS table. The radius and mass of maximum neutron stars are not so sensitive to the symmetry energy in these family TM1 parameterizations. However, the radii at intermediate mass region are strongly correlated with the slope of symmetry energy. Furthermore, the dimensionless tidal deformabilities of neutron stars are also calculated within the associated Love number, which is related to the quadrupole deformation of the star in a static external tidal field and can be extracted from the observation of gravitational wave generated by binary star merger. We find that its value at $1.4 M_\odot$ has a linear correlation to the slope of symmetry energy being different from the previous studied. With the latest constraints of tidal deformabilities from GW170817 event, the slope of symmetry energy at nuclear saturation density should be smaller than $60$ MeV in the family TM1 parameterizations. This fact supports the usage of lower symmetry energy slope for the update supernova EoS, which is applicable to simulations of neutron star merger. {Furthermore, the analogous analysis are also done within the family IUFSU parameter sets. It is found that the correlations between the symmetry energy slope with the radius and tidal deformability at $1.4 M_\odot$ have very similar linear relations in these RMF models. }

\end{abstract}



\maketitle

\section{Introduction}
%------------------------------------
The neutron star, as a super compact object in the universe,  is a fantastic natural laboratory to investigate the nuclear physics theory at extreme conditions, such as high density, high pressure, and high temperature~\cite{lattimer2016,oertel2017}. It is a possible product of supernova explosion and may be composed of neutrons, protons, leptons, and other hadrons with the strangeness degree of freedom. In the past half century, thousands of neutron stars were detected through various astronomical measurements~\cite{lattimer2005}. Their masses are around $1.2-2.0M_\odot$ and the radii are close to $10$ km~\cite{lattimer2005,martinez2015}. The discoveries of massive neutron stars ($2M_\odot$) provided an enormous constraint for the equation of state (EoS) of nuclear matter at high density~\cite{antoniadis2013,demorest2010,fonseca2016}, which plays a crucial role in the properties of neutron star through Tolman-Oppenheimer-Volkoff (TOV) equation~\cite{oppenheimer39,tolman39}.

Furthermore, in 2017,  the gravitational wave from merger of binary neutron stars was detected for the first time
(GW170817) by Advanced LIGO and Virgo collaboration~\cite{abbott2017a}. Soon afterwards, the short $\gamma$-ray burst (GRB170817A) and electromagnetic waves from X-ray to radio bands (AT2017gfo) were also measured~\cite{abbott2017b,goldstein17}.  These events activated the new era of multi-messenger astronomy and provided new measurements to observe the properties of neutron star. The tidal deformability extracted from the GW170817 event gave an additional constraint to neutron star properties, besides its mass and radius~\cite{abbott2017a}. It represents the quadrupole deformation of a neutron star due to a quadrupolar gravitational field from the companion star, which is related to the relativistic dimensionless Love number in the post-Newtonian expansion of the inspiral dynamics~\cite{mora2004}. The initial analysis from GW170817 data with a low-spin prior predicted $\tilde \Lambda\leq 800$, where $\tilde \Lambda$ denotes the combined dimensionless tidal deformability of the binary neutron star system. Actually, this constraint is very hard to estimate exactly and strongly model-dependent~\cite{abbott2018}.

Before GW170817 event, there were already several investigations to study the tidal deformability of neutron star and quark star with different EoSs which were assumed as polytropic form, or were obtained from various nuclear many-body methods, like Skyrme-Hartree-Fock (SHF) method,  variational method, relativistic mean-field (RMF) method, and so on~\cite{hinderer2008,hinderer2010,postnikov10,fattoyev2013,kumar2017,moustakidis2017}.  After August, 2017, more theoretical works were proposed to discuss the relations between the tidal deformability of neutrons star and EoS in such aspects as nuclear many-body methods, nucleon-nucleon interactions, phase transition, symmetry energy~\cite{annala2018,lim2018,most2018,fattoyev2018,kumar2018,malik2018,paschalidis2018,tews2018,zhang2018a,zhao2018,zhou2018,zhu2018}, etc.

The neutron star is composed of neutron-rich matter, where the symmetry energy $E_\text{sym}$ from the isospin effect of nucleons plays a very important role in determining the behaviors of EoS at high density and the structure of neutron star~\cite{danielewicz2002,danielewicz2014,li2008}. Its density-dependent tendency, the slope of symmetry energy $L$, is also a significant quantity to study the isospin difference of nuclear system and is strongly related to the neutron skin thickness of finite nuclei~\cite{rocamaza2011}. However, there is still no efficient method to restrict the symmetry energy at high density until now. On the other hand, the information of tidal deformability from the binary neutron star can provide a powerful tool to obtain a reasonable density-dependent behavior of symmetry energy.  In 2013, Fattoyev  et al. calculated the tidal polarizability of neutron star with different RMF interactions and found that the tidal polarizabilities were very sensitive to the symmetry energy at high density~\cite{fattoyev2013}. Recently, the neutron skin thickness of $^{208}$Pb was also constrained by the tidal deformation~\cite{fattoyev2018}.  Furthermore, Malik et al. used 18 RMF parameter sets and 24 SHF parameter sets, which can reproduce the ground-state properties of finite nuclei very well and support the existence of neutron stars with $2M_{\odot}$, to examine the correlations between tidal deformability and various nuclear saturation properties~\cite{malik2018}.

The crust of neutron star, as the non-uniform matter, is also strongly influenced by the symmetry energy and its slope. To discuss the effect of symmetry energy slope on the crust structure, we constructed parameter sets based on the TM1 interaction of RMF model~\cite{bao2014b}. {In fact, the TM1 interaction has been widely applied to the studies  of nuclear many-body systems and astrophysics, which achieved great successes due to its high density behaviors well constrained by relativistic Brueckner Hartree-Fock model. It also describes the properties of finite nuclei and the isoscalar properties of nuclear matter very well~\cite{sugahara94}. However, the TM1 interaction predicts rather large symmetry energy, which seems to be inconsistent with current constraints. This implies that the isovector part of the original TM1 parameterization need to be improved, which can be realized by introducing some higher order couplings among the mesons (see, for instance, Refs. \cite{dutra2014,horowitz2001}).} In these family TM1 parameterizations, the isoscalar saturation properties, e. g., saturation density, binding energy, and incompressibility, were kept and the symmetry energy was fixed at subsaturation density, $\rho_B=0.11$ fm$^{-3}$, while the slopes of symmetry energy at saturation density were chosen as different values from $40$ MeV to $110.8$ MeV through introducing an additional coupling term between $\omega$ meson and $\rho$ meson. In this case, the ground-state properties of finite nuclei are almost unchanged for different $L$. With the family TM1 interactions, the neutron drip density, the composition of the crust, and the phase transition of pasta phase were found to be strongly correlated with the slope of symmetry energy~\cite{bao2015}.

The present study is aimed to possible improvement of the EoS table for core-collapse
supernova simulations, which covers wide ranges of temperature, proton fraction, and baryon
density (see Table 1 in Ref. \cite{shen2011}). It is a difficult task to construct
a complete EoS over the whole range, since various phases exist at different thermodynamic
conditions and the description of phase transitions is very complicated.
Therefore, available EoSs for numerical simulations of core-collapse supernovae
are quite limited so far. The most commonly used EoSs in supernova simulations
are the Lattimer--Swesty EoS~\cite{lat91}, which employed
a compressible liquid-drop model with Skyrme force,
and the Shen EoS~\cite{shen1998a,shen1998b,shen2011}, which used a Thomas--Fermi approximation
based on the RMF model with the TM1 parameterization.
Both EoSs adopted the so-called single nucleus approximation (SNA), in which only a single
representative nucleus is considered instead of the distribution of nuclei.
Recently, a few EoS tables were developed beyond the SNA by considering the
full distribution of nuclei in nuclear statistical equilibrium using several
RMF parameterizations~\cite{hem10,fur11,fur17a,ste13,oertel2017}.

It is important for astrophysics community to provide the basic information of the family TM1 parameter set since it has been routinely used for many astrophysical simulations as the benchmark. In the last two decades, the Shen EoS table has been widely adopted in astrophysical simulations, such as core-collapse supernovae, proto-neutron star cooling, and black hole formation~\cite{sum07,nak13a}. It has been used for the construction of the supernova neutrino database for the prediction of supernova neutrino burst events and diffuse supernova neutrino background by Nakazato and Horiuchi.  
Furthermore, several extended versions of the Shen EoS table were developed by
including extra degree of freedoms as pions, hyperons, and quarks at high densities~\cite{ish08,nak08a,nak12}.
In the Shen EoS, both uniform matter and nonuniform matter employ the same nuclear
interaction (TM1), so that resulting thermodynamic quantities are consistent
and smooth in the whole range, which is important for performing numerical simulations.

It is now the transitional era for the widely used supernova EoS table with the TM1 parameter set due to the new observations. It has been noticed that the symmetry energy and its slope in the TM1 model
seem to be too large compared with the current constraints from neutron star observations.
The EoS with the TM1 interaction provides rather large neutron star radius,
which is inconsistent with the extracted value from X-ray observations~\cite{ste13,oertel2017}.
Furthermore, the detection of gravitational wave from neutron star merger GW170817
provides new constrains on the tidal deformability, which is also related to
a small neutron star radius.
It is well known that there exists a positive correlation between the neutron star radius
and the symmetry energy slope $L$~\cite{lattimer2014,Alam2016}.
By choosing a modified version of the TM1 model with a small value of $L$, it is suitable
to improve the description of neutron stars to be compatible with
observational constraints. Meanwhile, the properties of finite nuclei can be
satisfactorily reproduced as in the case of the original TM1 model.
Therefore, the extended TM1 model  with a smaller $L$ is considered
as a candidate for improving the Shen EoS table and keeping good descriptions of nuclei. To guide the choice of the slope parameter $L$, we intend to calculate the properties of neutron stars using the family TM1 parameterizations and compare the predictions with recent observations. This analysis of change in the family TM1 parameter sets is important for the reevaluation of the large scale simulations performed so far and for the prediction of new simulations of supernovae and neutron star mergers with the improved interaction.  Furthermore, we perform a similar calculation using the family IUFSU parameterizations so as to check the model dependence of our results.

In this paper, various properties of neutron stars, especially the tidal deformability will be calculated using two family parameterizations with different slopes of symmetry energy. The correlations between the slope of symmetry energy and the properties of neutron star will be examined with the same isoscalar saturation properties. This paper is arranged as follows. In section 2, the fundamental formulas will be shown. The numerical results and essential discussions will be given in section 3. Finally, the conclusion will be presented in section 4.

\section{The relativistic mean-field model for neutron star and tidal deformation}
The core region of neutron star can be considered as the uniform nuclear matter composed of mostly neutrons with a slight mixture of protons and leptons.  In the supernova core, hot and dense matter is a mixture of neutrons, protons, leptons, photons, and nuclei with uniform and non-uniform distributions.  In order to evaluate the thermodynamical properties of supernova matter, the RMF model at finite temperature is utilized \cite{shen1998b,shen2011}. 
In this work, we concentrate on the neutron star matter to provide the basic information and prepare for its applications to supernovae and neutron star merger.  Although there is a possibility to have hyperons or quark degrees of freedom, 
we study here in the RMF model without strangeness particles.  The RMF model describes the effective nucleon-nucleon interaction through exchanging the intermediate mass mesons, like $\sigma,~\omega$, and $\rho$ mesons. The RMF model has already achieved a lot of successes to investigate the properties of infinite nuclear matter and finite nuclei. There were hundreds of RMF parameter sets obtained from different points of view~\cite{dutra2014}. Here, we would like to adopt the following RMF Lagrangian based on TM1 interaction~\cite{sugahara94}, which is used to construct the series of Shen EoS tables,
\beq\label{tm1}
\mathcal{L}&= & \bar{\psi} ( i\gamma_{\mu}\partial^{\mu}- M_N- g_{\sigma}\sigma- g_{\omega}\gamma_{\mu}\omega^{\mu}-\frac{g_\rho}{2}\tau^a\gamma_\mu\rho^{a \mu } ){\psi} \nn
&    &+   \frac{1}{2}\partial_{\mu}\sigma\partial^{\mu}\sigma- \frac{1}{2}{m_{\sigma}}^2\sigma^2-\frac{1}{3}g_2\sigma^3-\frac{1}{4}g_3\sigma^4\nn
&    &-   \frac{1}{4}W_{\mu\nu}W^{\mu\nu}+\frac{1}{2}{m_{\omega}}^2\omega_{\mu}\omega^{\mu}+\frac{1}{4}c_3(\omega_{\mu}\omega^{\mu})^2\nn
&    &-\frac{1}{4}R^a_{\mu\nu}R^{a\mu\nu}+\frac{1}{2}{m_{\rho}}^2\rho^a_{\mu}\rho^{a\mu}+\Lambda_V(g^2_\omega\omega_{\mu}\omega^{\mu})(g^2_\rho\rho^a_{\mu}\rho^{a\mu}),
\eeq
where
\beq
&    &W_{\mu\nu}=\partial_\mu\omega_\nu-\partial_\nu\omega_\mu,\nn
&    &R^a_{\mu\nu}=\partial_\mu\rho^a_\nu-\partial_\nu\rho^a_\mu,
\eeq
are the antisymmetric field tensors of $\omega$ and $\rho$ mesons. The coupling term between $\omega$ meson and $\rho$ meson is introduced to control the density-dependent behaviors of symmetry energy with different $\Lambda_V$ values, which was firstly proposed in RMF model by Horowitz and Piekarewicz~\cite{horowitz2001}. We are applying this extension to update the Shen EoS table~\cite{shen2019}. 

With the Euler-Lagrangian equation, the equations of motion of nucleon and mesons are obtained,
\beq
&&\left[i\gamma_{\mu}\partial^{\mu}-(M_N+g_\sigma \sigma)
-g_\omega\gamma^\mu\omega_\mu
-\frac{g_\rho}{2}\tau^a\gamma_\mu\rho^{a\mu}\right]\psi=0,\nn
&   &(\partial^\mu\partial_\mu+m^2_\sigma)\sigma+g_2\sigma^2+g_3\sigma^3=-g_\sigma\bar\psi\psi,\nn
&   &\partial^\mu W_{\mu\nu}+m^2_\omega\omega_\nu+c_3(\omega_\mu\omega^\mu)\omega_\nu+2\Lambda_Vg^2_\omega g^2_\rho\rho^a_{\mu}\rho^{a\mu}\omega_{\nu}=g_\omega\bar\psi\gamma_\nu\psi,\nn
&   &\partial^\mu R^a_{\mu\nu}+m^2_\rho\rho^a_\nu+2\Lambda_Vg^2_\omega g^2_\rho\omega_{\mu}\omega^{\mu}\rho^a_{\nu}=g_\rho\bar\psi\gamma_\nu\tau^a\psi.
\eeq
These equations can be solved self-consistently in terms of mean-field approximation and no-sea approximation~\cite{walecka1974,ring1996,meng2006}. Furthermore, in a uniform system, the spatial derivatives of nucleon and mesons must be vanished. There are only the time components of mesons due to the rotational invariance. The energy density and pressure are the most important input in the study of neutron star, which can be generated by the energy-momentum tensor~\cite{serot1986},
\beq
T_{\mu\nu}=-g_{\mu\nu}\mathcal{L}+\frac{\partial\phi_i}{\partial x^\nu}\frac{\partial\mathcal{L}}{\partial(\partial\phi_i/\partial x_\mu)},
\eeq
where $\phi_i$ denotes the nucleon and various mesons. Finally, the energy density can be written as~\citep{shen2002},
\beq
\varepsilon &=&\sum_{i=n,p}\frac{2}{(2\pi)^3}\int_{|\bm k|<k^{i}_F}d^3\bm k\sqrt{k^2+M^{*2}}+g_\omega\omega\sum_{i=n,p}\rho^{i}_B+g_\rho\rho(\rho^p_B-\rho^n_B)\nn
&&+\frac{1}{2}m^2_\sigma\sigma^2+\frac{1}{3}g_2\sigma^3+\frac{1}{4}g_3\sigma^4-\frac{1}{2}m^2_\omega\omega^2-\frac{1}{4}c_3\omega^4-\frac{1}{2}m^2_\rho\rho^2-\Lambda_Vg^2_\omega g^2_\rho\omega^2\rho^2.
\eeq
Here, the time components of $\omega$ and $\rho$ mesons are simply expressed as $\omega$ and $\rho$. The $\rho_B$ is the baryon density of nucleon. The corresponding pressure is
\beq
p&=&\sum_{i=n,p}\frac{2}{3(2\pi)^3}\int_{|\bm k|<k^{i}_F}d^3\bm k\frac{k^2}{\sqrt{k^2+M^{*2}}}-\frac{1}{2}m^2_\sigma\sigma^2-\frac{1}{3}g_2\sigma^3-\frac{1}{4}g_3\sigma^4\nn
&&+\frac{1}{2}m^2_\omega\omega^2+\frac{1}{4}c_3\omega^4+\frac{1}{2}m^2_\rho\rho^2+\Lambda_Vg^2_\omega g^2_\rho\omega^2\rho^2.
\eeq
The symmetry energy of nuclear matter is defined as,
\beq
E_\text{sym}(\rho_B)=\left.\frac{1}{2}\frac{\partial^2\varepsilon(\rho_B,\delta)/\rho_B}{\partial\delta^2}\right|_{\delta=0},
\eeq
where $\delta$ is the isospin asymmetry $\delta=(\rho^n_B-\rho^p_B)/(\rho^n_B+\rho^p_B)$. The symmetry energy can be derived as an analytic expression~\citep{dutra2014},
\beq
E_\text{sym}(\rho_B)=\frac{k^2_F}{6\sqrt{M^{*2}_N+k^2_F}}+\frac{g^2_\rho\rho_B}{8(m^2_\rho+2\Lambda_Vg^2_\omega g^2_\rho\omega^2)}.
\eeq
$k_F$ is the Fermi momentum of symmetric nuclear matter. Its slope is given as,
\beq
L=3\rho_B\left(\frac{\partial E_\text{sym}}{\partial \rho_B}\right).
\eeq

In this work, the core of neutron star is considered as the constituents of neutron, proton, electron, and muon, which should satisfy the charge neutrality and $\beta$ equilibrium to generate a stable structure. Their chemical potentials are constrained by the following equations~\cite{shen2002},
\beq\label{mueq}
\mu_p&=&\mu_n-\mu_e,\nn
\mu_\mu&=&\mu_e.
\eeq
The chemical potentials of nucleons and leptons are related to their Fermi surfaces at zero temperature,
\beq\label{cmq}
\mu_i&=&\sqrt{k^{i2}_F+M^{*2}_N}+g_\omega\omega+g_\rho\tau_3\rho,\nn
\mu_l&=&\sqrt{k^{l2}_F+m^{2}_l},
\eeq
where $i=n,~p$ and $l=e,~\mu$.
The charge neutrality requires that the proton density is equal to the one of leptons,
\beq\label{nr}
\rho_p=\rho_e+\rho_\mu.
\eeq
The pressure and energy density will be obtained as a function of nucleon density under the constraints of Eqs. (\ref{mueq}) and (\ref{nr}). They are put into the TOV equation proposed by Tolman, Oppenheimer, and Volkoff to derive the properties of neutron star~\cite{oppenheimer39,tolman39},
\beq\label{tov}
\frac{dP(r)}{dr}&=&-\frac{GM(r)\varepsilon(r)}{c^2r^{2}}\frac{\Big[1+\frac{P(r)}{\varepsilon(r)}\Big]\Big[1+\frac{4\pi r^{3}P(r)}{M(r)c^2}\Big]}
{1-\frac{2GM(r)}{c^2r}},\nn
\frac{dM(r)}{dr}&=&4\pi r^{2}\varepsilon(r)/c^2,
\eeq
where, $c$ is the light speed. $P(r)$ is the pressure at radius $r$ and $M(r)$ is the total mass inside a sphere of radius $r$ of neutron star.

The dimensionless tidal deformability of neutron star is defined as~\cite{hinderer2010},
\beq
\Lambda=\frac{2}{3}k_2C^{-5},
\eeq
where $C=GM/Rc^2$ is the compactness parameter. $R$ and $M$ are the neutron star radius and mass, respectively. The dimensionless quadrupole tidal Love number $k_2$ is given by
\beq
k_2&=&\frac{8C^5}{5}(1-2C)^2[2+2C(y_R-1)-y_R]\nn
&&\bigg\{2C[6-3y_R+3C(5y_R-8)]\nn
&&+4C^3[13-11y_R+C(3y_R-2)+2C^2(1+y_R)]\nn
&&+3(1-2C)^2[2-y_R+2C(y_R-1)]\ln(1-2C)\bigg\}^{-1}.
\eeq
The quantity $y_R$ is the value of a function $y(r)$ at neutron star radius $R$. The function $y(r)$ is the solution of a first-order differential equation for $y$~\cite{hinderer2008},
\beq\label{td}
r\frac{dy(r)}{dr}+y(r)^2+y(r)F(r)+r^2Q(r)=0,
\eeq
with the boundary condition $y(0)=2$. The functions $F(r)$ and $Q(r)$ are related to energy density, pressure, and neutron star mass,
\beq
F(r)=\bigg\{1-4\pi r^2 G[\varepsilon(r)-P(r)]/c^4 \bigg\}\left(1-\frac{2M(r)G}{rc^2}\right)^{-1},
\eeq
and
\beq
Q(r)&=&\frac{4\pi G}{c^4}\left[5\varepsilon(r)+9P(r)+\frac{\varepsilon(r)+P(r)}{\partial P(r)/\partial\varepsilon(r)}\right]\left(1-\frac{2M(r)G}{rc^2}\right)^{-1}\nn
&&-6\left(r^2-\frac{2rM(r)G}{c^2}\right)^{-1}-\frac{4M(r)^2G^2}{r^4c^4}\nn
&&\left(1+\frac{4\pi r^3P(r)}{M(r)c^2}\right)^2\left(1-\frac{2M(r)G}{rc^2}\right)^{-2}.
\eeq
On the other hand,  the speed of sound in dense matter, $v_s$ is relevant to the derivative of pressure, $P(r)$ with respect to energy density, $\varepsilon(r)$,
\beq
\frac{\partial P}{\partial\varepsilon}=\left(\frac{v_s}{c}\right)^2.
\eeq

\section{The results and discussions}
\begin{figure*}[htb]
	\centering
	\includegraphics[width=8cm]{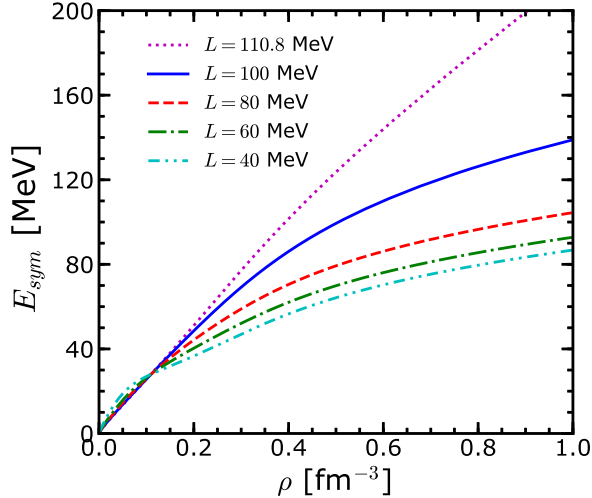}
	\caption{The symmetry energy as a function of density with different slopes, $L=40, ~60, ~80, ~100, ~110.8$ MeV for TM1 family parameter sets. }
	\label{esym}
\end{figure*}

To study the effect of symmetry energy on the properties of neutron-rich system, the family TM1 parameter sets were obtained by refitting the isovector coupling  constants in Lagrangian (\ref{tm1}), $g_\rho$ and $\Lambda_V$, with different slopes of symmetry energy at saturation density as $L=40,~50,~\dots,100$ MeV to compare with the original value from TM1, $L=110.8$ MeV. Another constraint is that all of the symmetry energies at a subsaturation density $\rho_B=0.11$ fm$^{-3}$ in these family sets were fixed as $E_\text{sym}=28.05$ MeV, which can minimize the influences on the binding energies of finite nuclei~\cite{bao2014b}. The symmetry energy as a function of density with different slopes from $40$ MeV to $110.8$ MeV are shown in Fig.~\ref{esym}. The smaller slope generates larger symmetry energy below the subsaturation density, while the situation is opposite at high density region, since the density-dependent behavior of symmetry energy can be approximately expressed by $E_\text{sym}(\rho)=E_\text{sym}(\rho_0)+L(\rho-\rho_0)/3\rho_0+\dots$ At high density region, the differences of symmetry energies with different slopes become very large, which will influence the neutron star properties obviously. The symmetry energy with $L=110.8$ MeV is about three times of the one with $L=40$ MeV at $\rho=0.8$ fm$^{-3}$.

\begin{figure*}[htb]
	\centering
	\includegraphics[width=8cm]{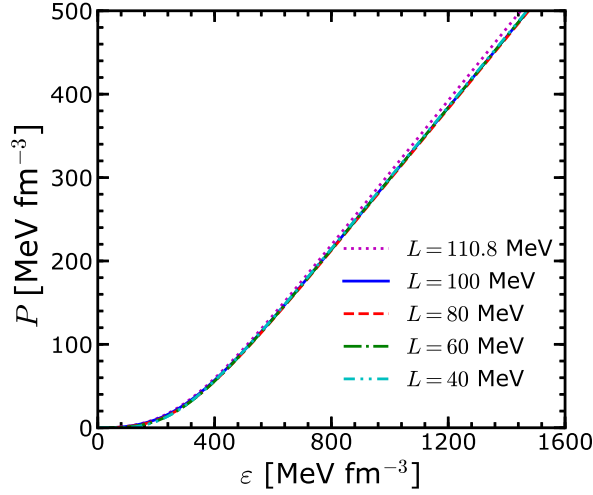}
	\caption{The equations of state of neutron star matter with different slopes, $L=40, ~60, ~80, ~100, ~110.8$ MeV for TM1 parameter set. }
	\label{eos}
\end{figure*}

The nuclear matter in the crust of neutron star has the non-uniform structure. In this work, a unified EoS at very low density region is taken from the Shen EoS at zero temperature, which was calculated within the Thomas-Fermi approximation in the framework of original TM1 parameter set~\cite{shen1998a,shen1998b,shen2011}. In this work, the effects of symmetry energy on neutron star properties are concentrated in discussing the uniform matter. The influences of symmetry energy including the non-uniform matter will be separately reported later.  The EoSs, $P(\epsilon)$ of neutron star matter in the core of neutron star can be obtained by solving the equations of motion of nucleon and mesons with the conditions of $\beta$ equilibrium and charge neutrality. The EoSs from family TM1 parameterizations are plotted in Fig.~\ref{eos} corresponding to different slopes of symmetry energy. In general, all EoSs look very similar. The larger $L$ generates a relatively stiffer EoS in detail.

\begin{figure*}[htb]
	\centering
	\includegraphics[width=8cm]{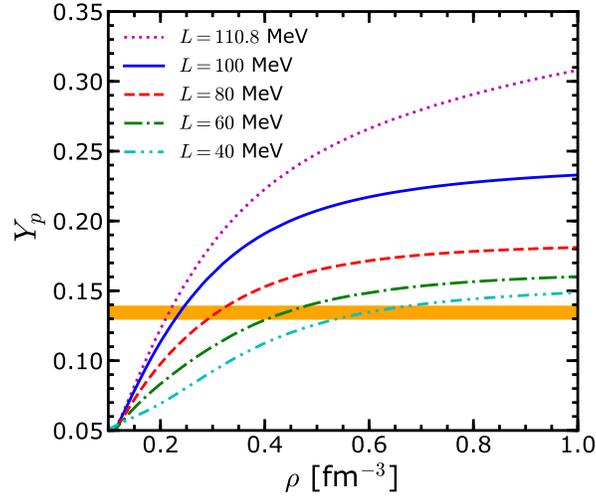}
	\caption{The proton fractions as functions of nucleon density in neutron star matter with different slopes, $L=40, ~60, ~80, ~100, ~110.8$ MeV for TM1 parameter set. The orange band represents the corresponding range of threshold values in DU process with different $L$.}
	\label{yp}
\end{figure*}

In Fig.~\ref{yp}, the proton fractions of uniform neutron star matter are shown as functions of nucleon density with family TM1 sets. They have the similar $L$-dependent behaviors with the symmetry energy. $Y_p$ increases with $L$. At high density, they become saturated due to the $\beta$-equilibrium. Actually, the proton fraction is approximately in proportion to the cube of symmetry energy for the free Fermi gas in neutron star~\cite{lattimer2016}. Furthermore, the magnitude of proton fraction is very important to determine whether the neutrino cooling process of neutron star, i. e. the direct Urca (DU) process occurs or not. Therefore, in the present framework, the threshold density of DU process is strongly dependent on the $L$. When the muon appears in neutron star, the threshold of DU process is also determined by the muon fraction due to the momentum conservation. In this work, the DU process happens starting from $Y_p=0.1298$ at $\rho=0.210$ fm$^{-3}$ for $L=110.8$ MeV, while this fraction will move to $Y_p=0.1389$ at $\rho=0.669$ fm$^{-3}$ for $L=40$ MeV.

\begin{figure*}[htb]
	\centering
	\includegraphics[width=8cm]{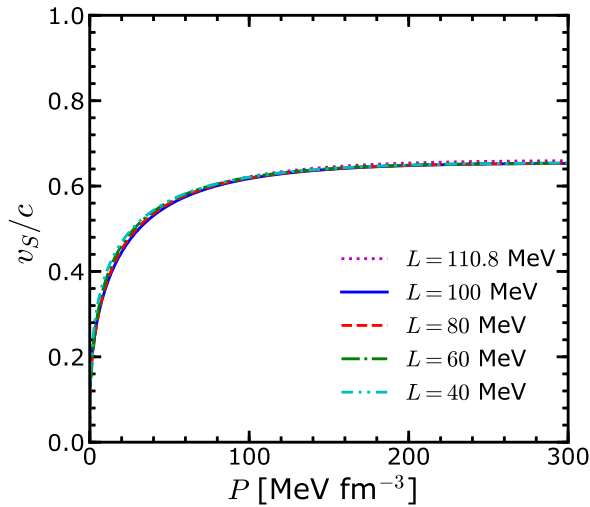}
	\caption{The speed of sound as a function of pressure in neutron star matter with different slopes, $L=40, ~60, ~80, ~100, ~110.8$ MeV for family TM1 parameter set. }
	\label{vs}
\end{figure*}

To obtain the Love number $k_2$, we need not only the EoSs but also the derivative of pressure respect to energy density, $\partial P/\partial \varepsilon$, which is related to the speed of sound in nuclear matter, $\partial P/\partial \varepsilon=(v_s/c)^2$. In Fig.~\ref{vs}, the speeds of sound in nuclear matter with different slopes of symmetry energy are shown as functions of pressure. At low density region, the speed of sound quickly increases and the EoS with smaller $L$ has higher $v_s$.  Its growth rate slows down as the density increases. Finally, it becomes saturated, whose value is around $0.6c$. It demonstrates that the pressure and energy density at high density has a linear relation approximately. Furthermore, our EoSs satisfy the requirement of relativity theory at high density, while many EoSs from the non-relativistic framework will lead to very large speed of sound at high density and even exceed the light speed~\cite{malik2018}.

\begin{figure*}[htb]
	\centering
	\includegraphics[width=8cm]{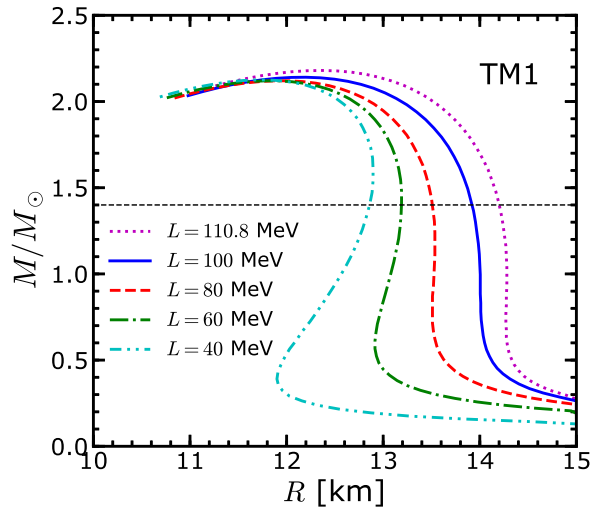}
	\caption{The neutron star mass-radius relations with different slopes, $L=40, ~60, ~80, ~100, ~110.8$ MeV for family TM1 parameter set. }
	\label{rmn}
\end{figure*}

\begin{figure*}[htb]
	\centering
	\includegraphics[width=8cm]{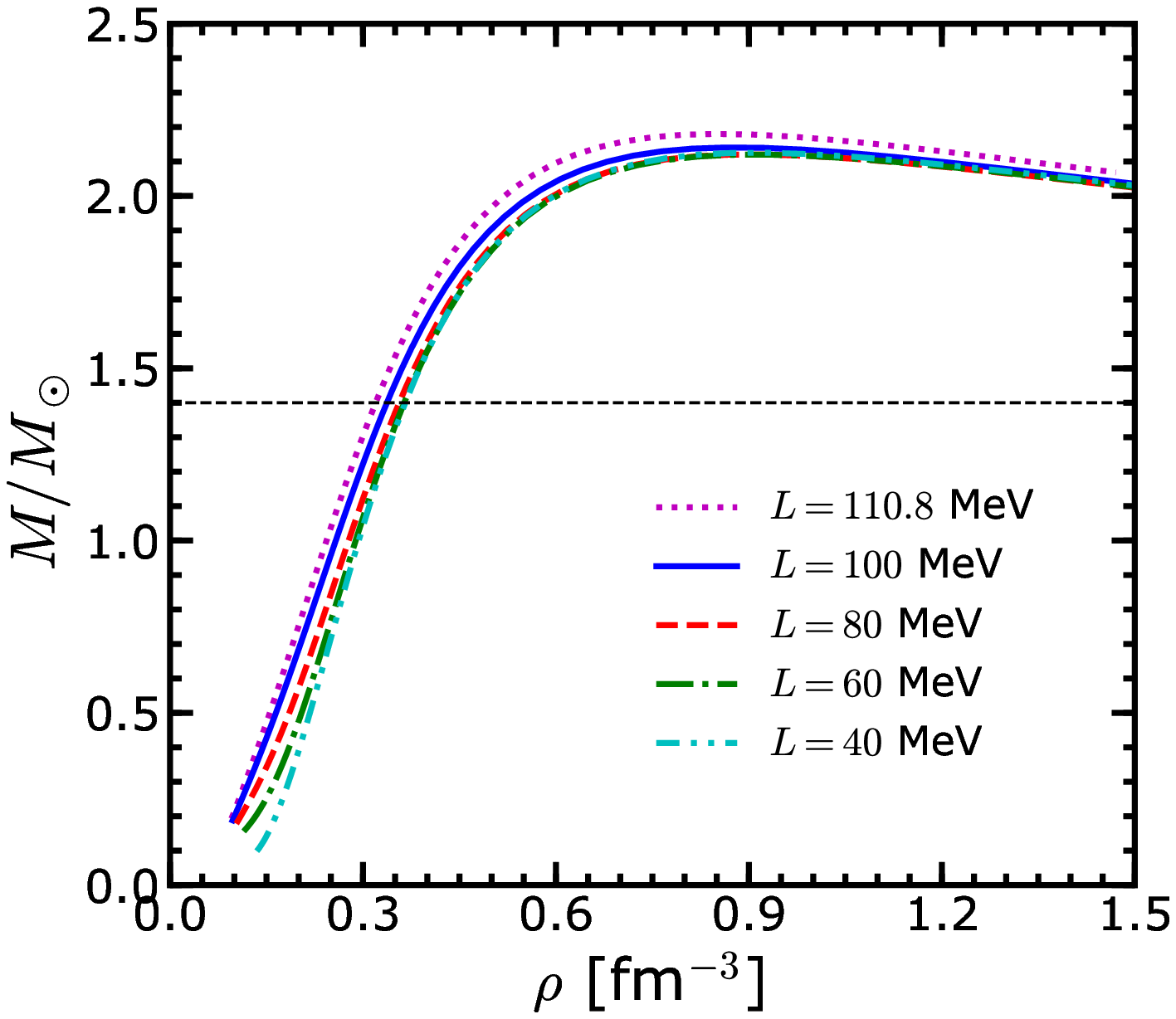}
	\caption{The neutron star mass-density relations with different slopes, $L=40, ~60, ~80, ~100, ~110.8$ MeV for family TM1 parameter set. }
	\label{rhom}
\end{figure*}

After solving the TOV equations with EoSs of neutron star matter, the properties of neutron star are obtained. The most important properties are the neutron star mass and its radius. In Fig.~\ref{rmn}, the neutron star mass-radius relations are given with different slopes of symmetry energy. It is found that the maximum masses of neutron star are not sensitive to the symmetry energy. They are in the range of $2.12-2.18 M_\odot$ for the family TM1 parameter sets with different slopes, $L$. On the other hand, the smaller $L$ can reduce the radius corresponding to the maximum mass slightly. It changes from $12.4$ km for $L=110.8$ MeV to $11.7$ km for $L=40$ MeV. However, the radius below $2M_\odot$ are significantly influenced by the slope of symmetry energy. In the past 50 years, the masses of neutron star were observed mainly in the range of $1.2M_\odot-2.0M_\odot$. Most of them are around $1.4M_\odot$. In the GW170817 event, the masses of binary neutron stars were also estimated around $1.4M_\odot$. Although the properties of neutron star with $1.4M_\odot$ have been important for the investigation, this evaluation of neutron stars attracts more attention in astrophysics and nuclear physics.  The radius of $1.4M_\odot$ neutron star is about $14.20$ km for the original TM1 parameter set with $L=110.8$ MeV. It decreases to $12.86$ km for $L=40$ MeV. The latest data analysis from LIGO and Virgo collaborations displayed the radii of binary neutron stars in GW170817 are $11.9\pm1.4$ km at the $90\%$ credible level~\cite{abbott2018}. Many other works also indicated $R_{1.4}$ should be smaller than $13.5$ km with the constraint of GW170817~\cite{de2018,most2018} in addition to the constraints from X-ray observations~\cite{fortin2015}.  The original TM1 parameter set with large $L$ used for the Shen EOS table is claimed to be excluded by the observational data of radii in the multi-messenger era, therefore, the revision of interaction for the improvement of the Shen EOS with a smaller $L$ is preferred based on the current results of family TM1 parameter set.  In Fig.~\ref{rhom}, the neutron star masses as functions of central density are plotted with different $L$. These curves are very similar to each other. The central densities corresponding to maximum neutron star mass are around $\rho=0.7-0.8$ fm$^{-3}$. They become about $0.3$  fm$^{-3}$ for $1.4M_\odot$ neutron star mass.

\begin{figure*}[htb]
	\centering
	\includegraphics[width=8cm]{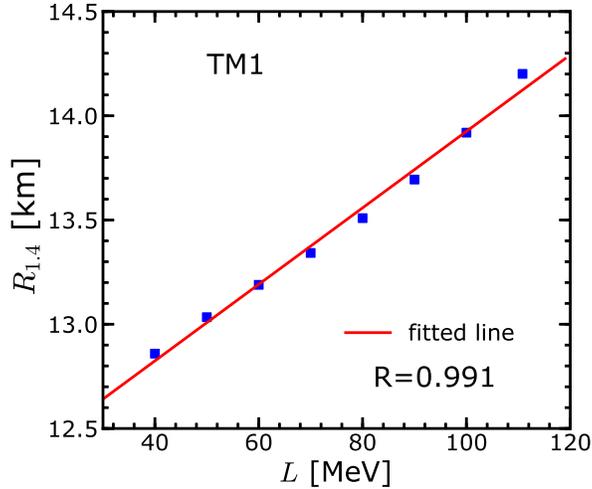}
	\caption{The relation between symmetry energy slope and neutron star radius at $1.4M_\odot$. }
	\label{rmr}
\end{figure*}

The relation between the radius of neutron star at $1.4M_\odot$, $R_{1.4}$, and the slope of symmetry energy in the family TM1 parameterizations is plotted in Fig.~\ref{rmr}. $R_{1.4}$ has a significant linear correlation with the slope of symmetry energy, $L$. The radius can be expressed as $R_{1.4}=12.091+0.0183L$ with a correlation coefficient, $R=0.991$. A larger $L$ results in a larger $R_{1.4}$ when we change $L$ in family TM1 parameter sets.  This is consistent with many other works~\cite{fattoyev2018,lim2018}.

\begin{figure*}[htb]
	\centering
	\includegraphics[width=8cm]{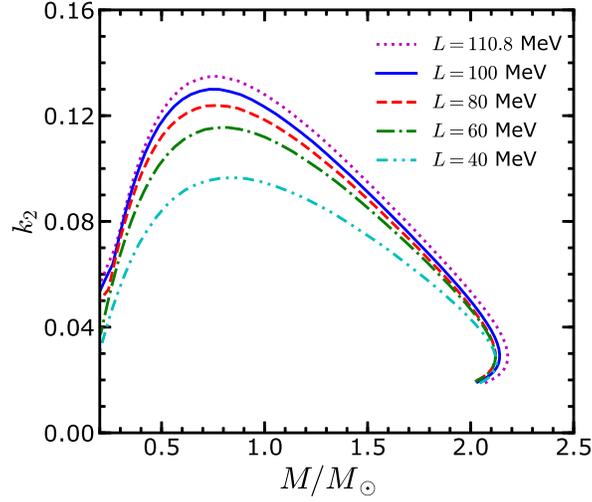}
	\caption{The Love number as a function of neutron star mass with different slopes, $L=40, ~60, ~80, ~100, ~110.8$ MeV for family TM1 parameter set. }
	\label{rmk}
\end{figure*}

\begin{figure*}[htb]
	\centering
	\includegraphics[width=8cm]{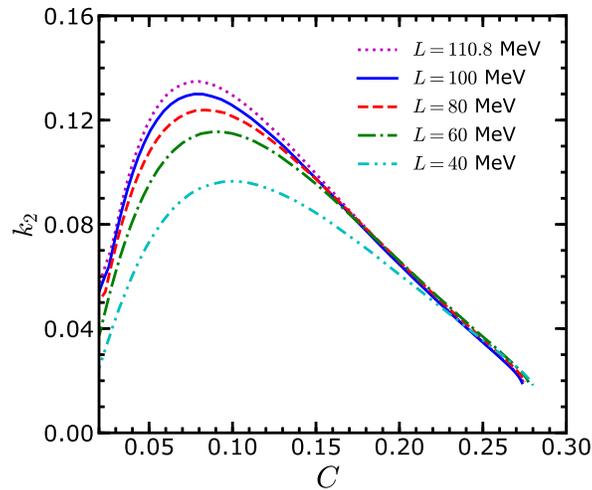}
	\caption{The Love number as a function of compactness parameter with different slopes, $L=40, ~60, ~80, ~100, ~110.8$ MeV for family TM1 parameter set. }
	\label{rmk1}
\end{figure*}

The Love number, $k_2$ is  obtained by solving the Eq. (\ref{td}), which is determined by the EoS of neutron star matter and its speed of sound. In Figs.~{\ref{rmk}} and {\ref{rmk1}}, the Love number is shown as a function of neutron star mass and compactness parameter, $C$, respectively with different slopes of symmetry energy. The Love number increases at small neutron star mass, arrives at the maximum value around $0.7-0.8M_\odot$, and  reduces rapidly in the larger mass region.  Its maximum value is obviously $L$-dependent, which is about $0.13$ for the original TM1 set with $L=110.8$ MeV and becomes $0.07$ in the case of $L=40$ MeV, while the Love numbers are very small and quite similar with different $L$ corresponding to smaller or larger neutron star masses. In general, the compactness parameter $C$ increases with neutron star mass, however, $y_R$ has opposite trend from the numerical calculation. Therefore, the behavior of $k_2$ is determined by the competition between $C$ and $y_R$.

\begin{figure*}[htb]
	\centering
	\includegraphics[width=8cm]{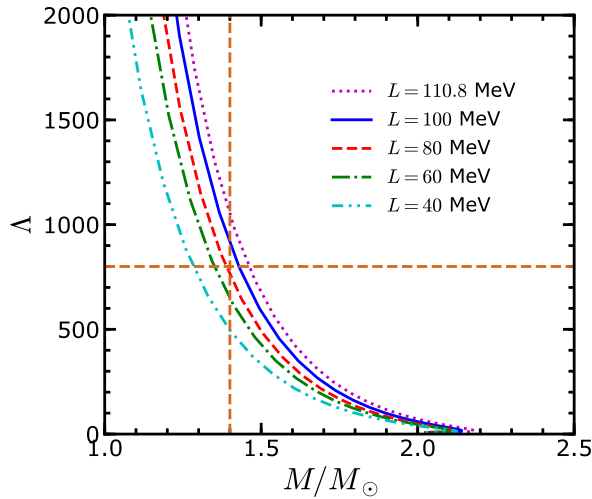}
	\caption{The tidal deformability as a function of neutron star mass with different slopes, $L=40, ~60, ~80, ~100, ~110.8$ MeV for family TM1 parameter set. The vertical dashed line denotes the neutron star mass, $1.4M_\odot$, while the horizontal one represents the tidal deformability $\Lambda=800$. }
	\label{rml}
\end{figure*}

The dimensionless tidal deformability becomes a particularly important quantity, which can be extracted by the detection of gravitational wave from binary neutron star merger. It provides another constraint on the EoS at high density besides the neutron star mass and radius. The dimensionless tidal deformabilities with different slopes of symmetry energy are given in Fig.~\ref{rml} as functions of neutron star masses. The tidal deformability of neutron star at small mass is very large, since $\Lambda\propto C^{-5}$, where the compactness parameter $C$ is very small. With neutron star mass increasing, it reduces to zero quickly. The tidal deformability at $1.4M_\odot$ are $496$ for $L=40$ MeV and $1045$ for $L=110.8$ MeV. Many analysis for the data of GW170817 event pointed $\Lambda_{1.4}<800$~\cite{abbott2017a,fattoyev2018}, which corresponds to $L=80$ MeV in this work.

\begin{figure*}[htb]
	\centering
	\includegraphics[width=8cm]{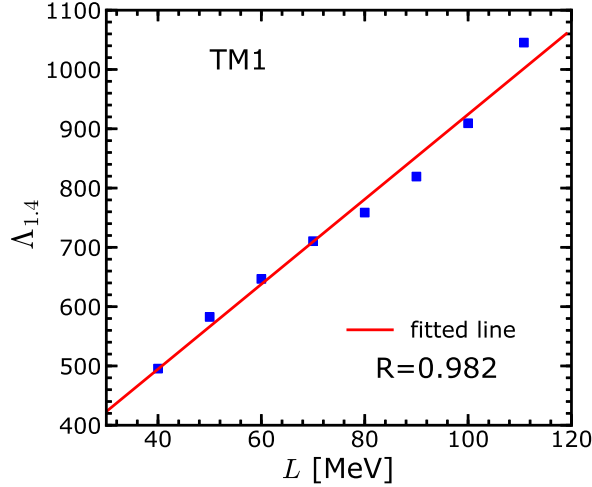}
	\caption{The relation between  symmetry energy slope and tidal deformability at $1.4M_\odot$. }
	\label{rm1}
\end{figure*}

In Fig.~\ref{rm1}, the relation between tidal deformability at $1.4M_\odot$  and the slope of symmetry energy, $L$ is shown within the family TM1 parameter sets. There is also a linear correlation between them, where $\Lambda_{1.4}=208.23+7.16L$ with a correlation coefficient $R=0.982$.  Actually, in the work of Lim and Holt, the Bayesian analysis method also supported an analogous conclusion with the EoSs from the chiral effective field theory~\cite{lim2018}.

\begin{figure*}[htb]
	\centering
	\includegraphics[width=8cm]{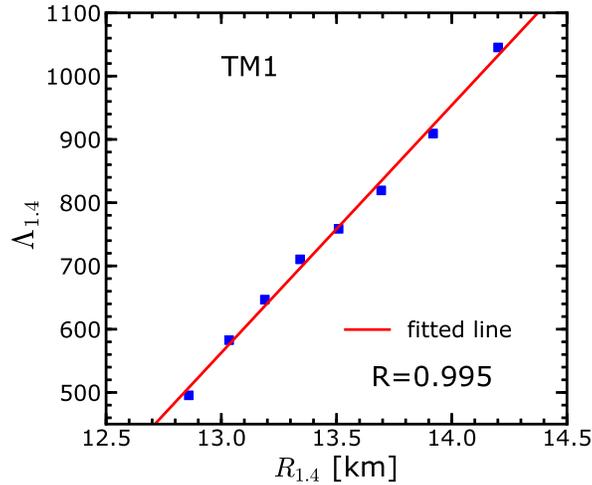}
	\caption{The relation between neutron star radius and tidal deformability at $1.4M_\odot$. }
	\label{rm}
\end{figure*}

Similarly, the $\Lambda_{1.4}$ is linearly dependent on the radius of neutron star at $1.4 M_{\odot}$, $R_{1.4}$, plotted in Fig.~\ref{rm}, because of the linear correlation between $R_{1.4}$ and $L$ in Fig.~\ref{rmr}.  They have a highly linear correlation comparing to $R_{1.4}$-$L$ and $\Lambda_{1.4}$-$L$ with $\Lambda_{1.4}=-4522.34+391.16R_{1.4}$, where the correlation coefficient is $0.995$. However, in many conventional works~\cite{fattoyev2018,malik2018}, the correlation between $R_{1.4}$ and $\Lambda_{1.4}$  was considered as a power exponent function. Note that those works have been made using a collection of various interactions with different forms. Therefore, it is necessary to discuss the properties of neutron star in a systematic manner under controlling. This relation with $\Lambda_{1.4}<800$ leads to a constraint on neutron star radii $R_{1.4}< 13.6$ km in the models for family TM1 parameter set.

\begin{figure*}[htb]
	\centering
	\includegraphics[width=8cm]{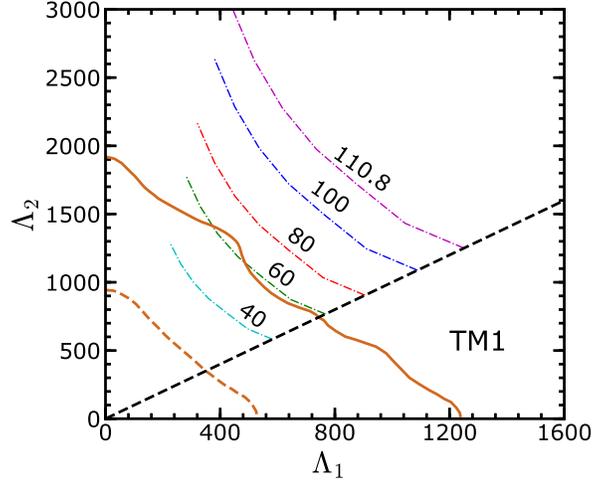}
	\caption{The tidal deformabilities of the two binary components of GW170817 for different symmetry energy slopes and, $50\%$ and $90\%$ credible levels extracted from the gravitational-wave detectors. }
	\label{rmll}
\end{figure*}

The chirp mass $\mathcal{M}=(m_1m_2)^{3/5}(m_1+m_2)^{-1/5}$ for GW170817 event, was observed as $1.188M_\odot$, where $m_1$ and $m_2$ are the masses of binary neutron star merger, respectively~\cite{abbott2017a}. The precise values of $m_1$ and $m_2$ cannot be determined solely by the gravitational-wave signal in this case. Therefore, a high-mass component, $m_1$ is assumed to be in the range from $1.365 M_\odot$ to $1.600 M_\odot$. The corresponding lower mass, $m_2$ should be from $1.170 M_\odot$ to $1.365 M_\odot$. The relevant tidal deformability $\Lambda_1(m_1)$ and $\Lambda_2(m_2)$ are plotted in Fig.~\ref{rmll} with different slopes of symmetry energy, $L$ with dot-dashed lines. The $50\%$ and $90\%$ credible levels of tidal deformability from the latest analysis of GW170817 event for the low-spin prior by LIGO and Virgo collaboration are denoted as dashed and solid curves, respectively~\cite{abbott2018}. The present $90\%$ confidence limit is lower than the one in the first analysis by LIGO and Virgo collaboration and prefers a softer EoS. With these constraints, the slope of symmetry energy, $L$ should be smaller than $60$ MeV in the TM1 family parameter set. The original TM1 set is excluded due to its larger $L$. However, the improved version of TM1 with $L=40$ MeV satisfies the constraints and can be used for supernovae and neutron star mergers.  

Implications of the allowed range of small value of $L$ are helpful to consider nuclear and astrophysical applications of the RMF model.  A small value of $L=40$ MeV, for example, means small symmetry energies at high densities.  This is consistent with the recent evaluation of symmetry energy at 2$\rho_0$ \cite{zhang2018b}.  The small value of $L$ leads to small proton fractions inside neutron stars and hinders a possibility of Direct Urca process as seen in Fig.~\ref{yp}.  The symmetry energy at sub-saturation density is slightly larger than that in the original TM1 case and the behavior of neutron matter becomes close to those in microscopic approaches.  The properties of inner crust and pasta phase of neutron stars are accordingly affected \cite{bao2014b,bao2014a}.  The neutron drip density becomes low and the appearance of pasta phase is enriched.  It is expected that different density-dependence of symmetry energy may affect the cooling of proto-neutron stars \cite{sum95,nakazato2019}.  

The systematic feature of neutron stars found in the family TM1 parameter sets provides us with the guidance to improve the supernova EoS.  The Shen EoS with the original TM1 parameter set has a large value of $L$ and leads to too large values for the neutron star radius and tidal deformability.  We can remedy these problems by choosing a value of $L$ smaller than 60 MeV.  Note that the behavior of symmetric nuclear matter is exactly the same and the maximum neutron star mass remains similar.  We recently constructed the table of EoS for supernova simulations with an improved TM1 parameter set with $L=40$ MeV \cite{shen2019}.  Astrophysical applications of the dense matter at finite temperature with the improved TM1 parameter set have been made~\cite{sum19} and will reported elsewhere.  

\begin{figure*}[htb]
	\centering
	\includegraphics[width=8cm]{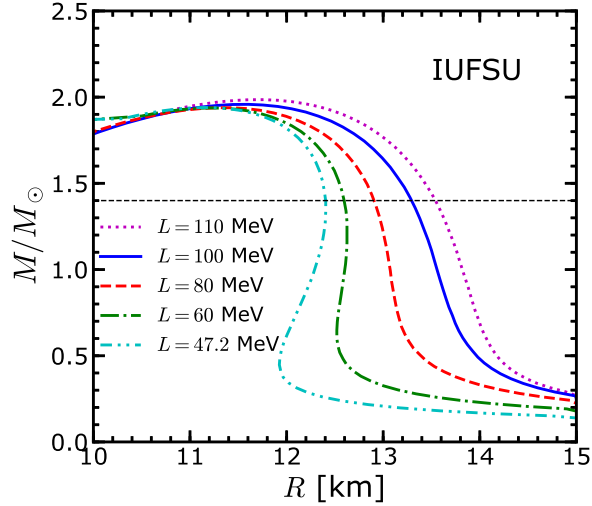}
	\caption{The neutron star mass-radius relations with different slopes, $L=47.2, ~60, ~80, ~100, ~110$ MeV for family IUFSU parameter sets. }
	\label{iufsurmn}
\end{figure*}

{In order to check the model dependence of the correlations obtained above, we perform a similar calculation using another family parameter sets, IUFSU. The original IUFSU model~\cite{fattoyev10} was proposed to overcome a smaller neutron-star mass predicted by the FSU model, and meanwhile it could keep an excellent description of ground-state properties of finite nuclei. In Ref.~\cite{bao2014b}, two family parameterizations, IUFSU and TM1, were generated by refitting the isovector coupling constants, $g_\rho$ and $\Lambda_V$, to achieve different density-dependent behaviors of symmetry energy. These family RMF parameterizations are useful for examining the correlations between the symmetry energy slope and neutron star properties. In Fig.~\ref{iufsurmn}, the neutron star mass-radius relations were displayed using the family IUFSU parameter sets with $L=47.2, ~60, ~80, ~100, ~110$ MeV, respectively. Note that the original IUFSU model predicts the value $L=47.2$ MeV. The variety of $L$ does not influence the maximum mass of neutron star and the corresponding radius too much, which is very similar to the results from family TM1 parameter sets as shown in Fig.~\ref{rmn}. The effect of $L$ mainly embodies in radii at the low mass region. A larger $L$ generates a larger radius at $1.4M_\odot$.} 

\begin{figure*}[htb]
	\centering
	\includegraphics[width=8cm]{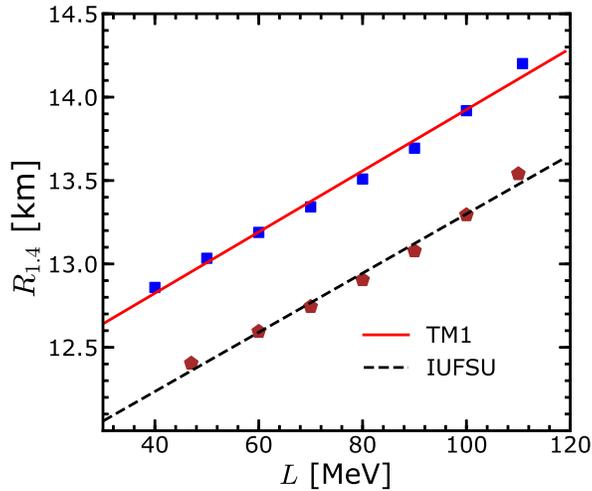}
	\caption{The relation between symmetry energy slope and neutron star radius at $1.4M_\odot$ in family IUFSU and TM1 parameter sets. }
	\label{iufsurmr}
\end{figure*}

\begin{figure*}[htb]
	\centering
	\includegraphics[width=8cm]{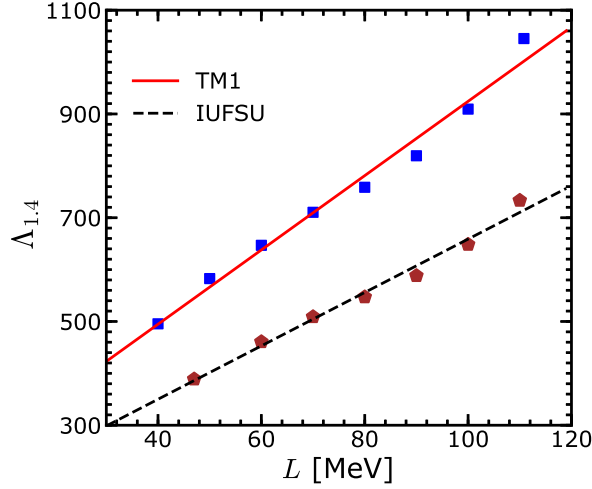}
	\caption{The relation between symmetry energy slope and tidal deformability at $1.4M_\odot$  in family IUFSU and TM1 parameter sets. }
	\label{iufsurm1}
\end{figure*}

\begin{figure*}[htb]
	\centering
	\includegraphics[width=8cm]{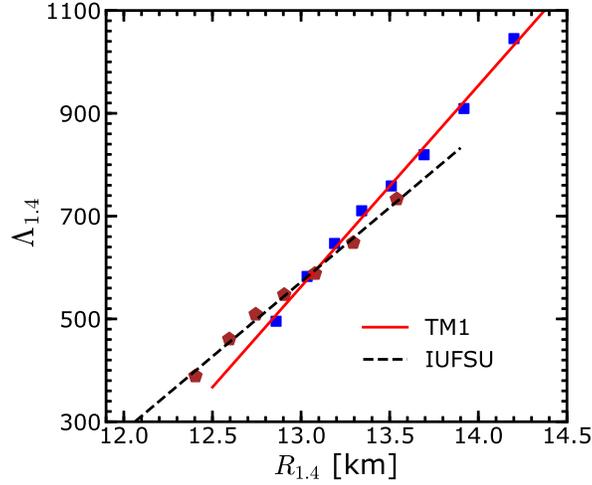}
	\caption{The relation between neutron star radius and tidal deformability at $1.4M_\odot$  in family IUFSU and TM1 parameter sets. }
	\label{iufsurm}
\end{figure*}

{In Figs.~\ref{iufsurmr}-\ref{iufsurm}, the relations about the slope of symmetry energy, $L$, the radius and tidal deformability of neutron star at $1.4M_{\odot}$, $R_{1.4}$-$L$, $\Lambda_{1.4}$-$L$, and $\Lambda_{1.4}$-$R_{1.4}$, from the family IUFSU parameter sets are plotted and are compared to those from TM1 parameter sets, respectively. All of them satisfy the linear relations very well, which are consistent with the conclusions from family TM1 parameter set as we discussed before. Here we emphasize again that these linear relations have the significant differences with other investigations relevant to $\Lambda_{1.4}$-$R_{1.4}$, where tidal deformability was written in a power function as, $\Lambda_{1.4}=aR^\alpha_{1.4}$~\cite{fattoyev2018,malik2018}. The possible reason is that they discussed this relation with a collection of different theoretical frameworks and model parameters. In that case, it is very difficult to control the invariance of isoscalar properties of nuclear matter. }

\begin{figure*}[htb]
	\centering
	\includegraphics[width=8cm]{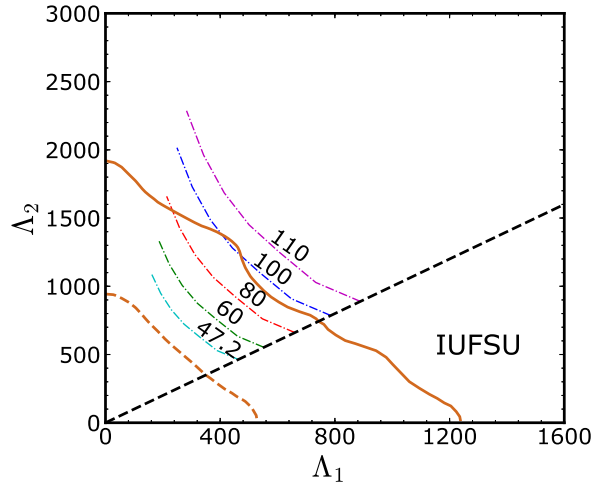}
	\caption{The tidal deformabilities of two binary components of GW170817 for different symmetry energy slopes from family IUFSU sets and, $50\%$ and $90\%$ credible levels extracted from the gravitational-wave detectors. }
	\label{iufsurmll}
\end{figure*}

{The tidal deformabilities from family IUFSU parameter sets are shown in Fig.~\ref{iufsurmll} together with the constraints from the GW170817 event. It can be found that the slopes of symmetry energy below $80$ MeV in family IUFSU sets are preferred by the gravitational-wave observations, which is also consistent with other constraints for $L=58.7\pm28.1$ MeV~\cite{oertel2017}. The different constraints from TM1 and IUFSU parameter sets are mainly caused by their incompressibilities, $K$. The EOSs of TM1 parameter sets with larger $K=281$ MeV are stiffer than those, $K=231$ MeV of IUFSU parameter sets, which result in the larger neutron star radii and larger tidal deformabilities.}

%-------------------------------------------------------------------------------
\section{Conclusions}
The properties of neutron star, such as mass, radius, and tidal deformability were investigated in the framework of relativistic mean-field (RMF) model. To study the effect of symmetry energy on neutron star, a coupling term between $\omega$ and $\rho$ mesons was introduced in the Lagrangian of RMF model with TM1 parameter set. The coupling constants related to $\rho$ meson were refitted by a fixed symmetry energy at sub-saturation density and different values of slope parameter $L$ of symmetry energy at saturation density, while the other coupling constants remain the same as original TM1 in order to keep the isoscalar saturation properties of nuclear matter satisfying the empirical data as well as good properties of nuclei. Therefore, the family TM1 parameter sets are suitable to assess different density-dependent behavior of symmetry energy.

This study is aimed to provide basic properties of neutron stars in the extended RMF model with the TM1 parameter set, which is used for the equation of state for supernova simulations.  Although the Shen EoS with the original TM1 parameter set has been widely used in astrophysical simulations, recent astronomical observations on neutron star radii are putting constraints on the symmetry energy, which was determined with the limited knowledge at the time of determination.  The symmetry energy in the original TM1 parameter set has been claimed too large and there is a room to improve the isovector part of interaction.  We would like to revise the TM1 parameter set for the improvement of supernova EoS while keeping good properties of nuclear matter.  Therefore, we explore variations of the slope of symmetry energy for the preparation of supernova EoS. The information of this analysis is used for astrophysical community to prepare new simulations with the family TM1 parameter sets.  

In this work, we discussed the symmetry energy effects on the properties of neutron star with the family TM1 parameter sets.  
The equations of state (EoSs) of neutron star matter with nucleon and lepton under $\beta$ equilibrium and charge neutrality conditions were taken into the TOV equation to obtain the properties of neutron star. The radii at intermediate mass region, like $1.4M_\odot$, are largely impacted by the change of the slope of symmetry energy in the RMF model. A smaller value of slope parameter $L$ generates a smaller radius of neutron star at $1.4M_\odot$. 
By changing the slope parameter in the family TM1 parameter sets, we can reduce the radius while keeping the maximum mass.  Since the neutron star radius of Shen EoS is claimed too large, the current study suggests that the modified TM1 with a small $L$ value is suitable for the improvement of supernova EoS. Furthermore, the radius $R_{1.4}$ is found to have a strong correlation with the slope of symmetry energy by using the family TM1 parameter sets in this work. This is well in accord with the previous studies showing the correlation in other frameworks of EoSs.  

The tidal deformability of neutron star is an important quantity in the binary neutron star merger, which can be extracted by the gravitational-wave detector.  We evaluated the tidal deformability by solving the first-order differential equation with the speed of sound in nuclear matter. The magnitude of tidal deformability, $\Lambda$, is found to have a strong dependence on the slope of symmetry energy. There is a linear correlation between the tidal deformability at $1.4M_\odot$, $\Lambda_{1.4}$ and the slope of symmetry energy. The $\Lambda_{1.4}$ and $R_{1.4}$ has a similar linear relation as well. With the constraint of latest data analysis from the GW170817, the slope of symmetry energy at nuclear saturation density, $L$ should be smaller than $60$ MeV in the family TM1 parameter sets from the tidal deformability of neutron star. This provides us with the guidance to adopt a small value of $L$ for the improvement of supernova EoS.  Accordingly, we will study the supernova matter at finite temperature with a revised TM1 interaction in the coming works \cite{shen2019} and its astrophysical applications \cite{sum19} including the neutron star merger. 

{To check the model dependence of the correlations obtained in the  family TM1 parameter sets, we also performed a similar calculation using the family IUFSU parameter sets. It was found that the correlations between the symmetry energy slope with the radius and tidal deformability at $1.4M_\odot$ have very similar linear relations as the case of TM1. }

There are caveats and extensions that need further studies in the current work.  In the neutron star models, 
the EoS of crust region was fixed with the one from the original TM1 set. It would be better to investigate neutron star properties with a unified EoS, where both of crust and core regions are described by the family TM1 parameterizations with different $L$.  The corresponding extension of Shen EoS table at finite temperature case and its applications to supernovae are also underway.

\section{Acknowledgments}
This work was supported in part by the National Natural Science Foundation of China (Grants  No. 11775119, No. 11675083, and No. 11405116), the Natural Science Foundation of Tianjin, and China Scholarship Council (Grant No. 201906205013 and No. 201906255002). This work is supported by Grant-in-Aid for Scientific Research (19K03837, 15K05093) and Grant-in-Aid for Scientific Research on Innovative areas
"Gravitational wave physics and astronomy:Genesis" (17H06357, 17H06365) and "Unraveling the history of the universe and matter evolution with underground physics" (19H05811) from the Ministry of Education, Culture, Sports, Science and Technology (MEXT), Japan.  
KS acknowledges high performance computing resources at KEK, JLDG, RCNP, YITP and UT. This work was partly supported by
the research programs at K-computer of the RIKEN AICS, HPCI Strategic Program of Japanese MEXT, “Priority Issue on Post-K computer” (Elucidation of the Fundamental Laws and Evolution of the Universe) and Joint Institute for Computational Fundamental Sciences (JICFus).

%\clearpage

\end{document}